\documentclass[twocolumn]{aastex631}

\usepackage{graphicx,times}             
\usepackage{natbib}
\usepackage{amssymb,amsmath}
\bibpunct{(}{)}{;}{a}{}{,}

\newcommand{\degree}{\hbox{$^\circ$}}

\newcommand{\s}{\ensuremath{\mbox{~s}}}

\newcommand{\cm}{\ensuremath{\mbox{~cm}}}
\newcommand{\pcmsq}{\ensuremath{\cm^{-2}}}

\newcommand{\vel}{km\,s$^{-1}$}

\newcommand{\um}{$\mu$m}

\newcommand{\egcite}{\citep[e.g.,][]}

\newcommand{\htcop}{H$^{13}$CO$^{+}$}
\newcommand{\sig}{$\sigma$}

\newcommand\avir{{$\alpha_{\rm vir}$}}

\newcommand\Lcal{{\cal{L}}}

\begin{document}

\title{Implication of the velocity dispersion scalings on high-mass star formation in molecular clouds}

\shorttitle{Velocity dispersion scalings linked to star formation}
\shortauthors{An-Xu Luo et.al}


\correspondingauthor{Hong-Li Liu}
\email{hongliliu2012@gmail.com}

\author{An-Xu Luo}

\affil{School of physics and astronomy, Yunnan University, Kunming, 650091, PR China \\}
\affil{Both authors contributed equally to this work.\\}

\author{Hong-Li Liu}
\affil{School of physics and astronomy, Yunnan University, Kunming, 650091, PR China \\}
\affil{Both authors contributed equally to this work.\\}

\author{Sheng-Li Qin}
\affiliation{School of physics and astronomy, Yunnan University, Kunming, 650091, PR China \\}

\author{Dong-ting Yang}
\affiliation{School of physics and astronomy, Yunnan University, Kunming, 650091, PR China \\}

\author{Sirong Pan}
\affiliation{School of physics and astronomy, Yunnan University, Kunming, 650091, PR China \\}

\begin{abstract}

This paper is aimed at exploring implications of velocity dispersion scalings on high-mass star formation in molecular clouds, including the scalings of Larson's linewidth--size ($\sigma$--$R$) and ratio--mass surface density ($\Lcal$--$\Sigma$; here ${\Lcal}$$=\sigma/R^{0.5}$). We have systematically analyzed the $\sigma$ parameter of well-selected 221 massive clumps, complemented with published samples of other hierarchical density structures of molecular clouds over spatial scales of 0.01--10\,pc. Those massive clumps are classified into four phases: quiescent, protostellar, HII region, and PDR clumps in an evolutionary sequence. The velocity dispersion of clumps increases overall with the evolutionary sequence, reflecting  enhanced stellar feedback in more evolved phases. The relations of $\sigma$--$R$ and $\Lcal$--$\Sigma$ are weak with the clump sample alone, but become evident when combined with others spanning a much wider spatial scales. For $\sigma$--$R$, its tight relation indicates a  kinematic connection between hierarchical density structures, supporting theoretical models of multiscale high-mass star formation. From the $\Lcal$--$\Sigma$ relation, cloud structures can be found to transition from over-virial state ($\alpha_\mathrm{vir} > 2$) to sub-virial state ($\alpha_\mathrm{vir} < 2$) as they become smaller and denser, indicating a possible shift in the governing force from turbulence to gravity. This implies that the multiscale physical process of high-mass star formation hinges on the self-gravity of sub-virial molecular clouds. However, the influence of turbulence may not be dismissed until large-scale clouds attain a sub-virial state. This is pending confirmation from future multiscale kinematic observations of molecular clouds with uniform observing settings.
\end{abstract}


\section{Introduction} \label{sec:intro}
Stars form in molecular clouds (MCs) that consist of hierarchical density structures, from filaments, clumps,  cores, and down to seeds of star formation. Formation of high-mass stars has been far from being understood compared to low-mass stars \citep{Mot18}. Recently, it is increasingly accepted that high-mass star formation could be a multiscale process that proceeds on all hierarchical density structures of MCs in a top-down manner, from filaments down to seeds of star formation \egcite{Per13, Mot18, Vaz19, Liu22a, Liu22b, Liu23, Yang23}. This multiscale phenomenon could be regulated by several basic factors such as turbulence, gravity, and magnetic field of MCs.

The turbulence in MCs could drive the multiscale physical process related to high-mass star formation. Different hierarchical density structures of MCs could be kinematically correlated through coherent motions across a wide range of scales, from 0.1\,pc up to 100\,pc, as reflected from the Larson's linewidth--size relation \egcite{Larson1981,Solomon1987}, which is one of the empirical velocity dispersion scalings. In theory, supersonic turbulence can generate an intricate network of filaments \egcite{Ino13,Gong15}, as well as denser substructures (e.g., clumps/cores) with non-linear density fluctuations, which are the sites of star formation \egcite{Elm1993, Bal11b, Kri11, Hen12, Pan19, Pad20}. This implies a connection between supersonic turbulence and star formation. This connection has been strengthened by simulations of MCs with supersonic turbulence triggered by supernova explosions \citep{Pad16}. The simulations reproduced the power-law form of the Larson's velocity-size relation, that is $\sigma \propto\mathit{R}^{0.39}$, where $\sigma$ is the velocity dispersion. In this context, \cite{Pad20} posited  high-mass stars form in MCs by a multiscale mass accretion/transfer process from large-scale converging inertial flows through clumps and cores to seeds of star formation, where large-scale inertial flows are thought to originate from  supersonic turbulence.

In addition to turbulence, gravity could regulate the multiscale dynamical scenario of high-mass star formation. The ``global hierarchical collapse" (GHC) model proposed that the multi-scale accretion process in high-mass star formation is primarily governed by gravity on all scales \citep{Vaz19}. In this model, MCs evolve by hierarchical and chaotic gravitational collapse from large-scale clouds through clumps and cores down to seeds of star formation \egcite{Har07, Gon14, Vaz17}. This gravitational collapse is accompanied by the release of gravitational energy, which in turn generates kinetic energy \egcite{Vaz09, Bal11a, Vaz19, Bal20}. As suggested in \citet{Vaz19}, this energy conversion could be reflected from the Larson's ratio--mass surface density ($\Lcal$--$\Sigma$) relation (see below for more discussions),  another velocity dispersion scaling, where ${\Lcal}$ is define as $\sigma/R^{0.5}$. Moreover, the global hierarchical collapse of cloud structures was claimed to facilitate formation of high-mass stars. 

Recent observations have provided evidence for a multiscale scenario of high-mass star formation, highlighting the significance of both turbulence and gravity in this process \citep{Liu22a,Liu22b, Liu23, Zhou22, Yang23, Pan23}. For instance, \citet{Yang23} and \citet{Pan23} observed a multiscale, dynamical mass accretion process in two different hub-filament system clouds (i.e., G310 and G34). Such observational evidence supports the latest generation of GHC and I2 models, but they differ in driving mechanisms for this scenario, particularly on the larger scales. The GHC model supports a gravity-driven hierarchical process, whereas the I2 model suggests a turbulence-driven mechanism. This disparity was also observed in previous investigations of a filamentary infrared dark cloud, G34, at $\sim 2$ arcsecond resolution \egcite{Liu20, Liu22a, Liu22b}. Those authors argued that the scale-dependent combined effect of turbulence and gravity is crucial for high-mass star formation in the cloud. Therefore, a comprehensive observational study of the interaction between turbulence and gravity in a large sample of cloud structures is necessary. As mentioned earlier, the velocity dispersion scalings would be an useful approach for such study.

This paper presents a study primarily on the gas kinematics (i.e., velocity dispersion) of 221 massive star-forming clumps. The \htcop~(1--0) molecular line is used as a tracer of the velocity dispersion. The goal is to understand the multiscale high-mass star formation scenario from the velocity dispersion scaling across all hierarchical density structures of MCs. To this end, our data set of massive clumps are complemented with samples of other hierarchical density structures from the literature (see below). The paper is structured as follows: Section\,2 describes the main data sets used in this study; Section\,3 presents the results and analysis; Section\,4 discusses the main findings; and Section\,5 summarizes the conclusions.

\section{Data set} \label{sec:datasets}
The Millimetre Astronomy Legacy Team 90\,GHz (MALT90) survey is a molecular line survey at  90\,GHz that aims to characterize the physical and chemical evolution of massive star-forming clumps \citep{Jac13}. This survey observed a total of 3,246 massive star formation clumps using the Mopra 22m-aperture single-dish telescope at an angular resolution of $\sim 38^{\prime\prime}$ and a sensitivity of 0.25\,K for the 0.11\,{\vel} velocity resolution. The targeted clumps are a subsample of the sources observed by the ATLASGAL 870\,$\mu$m dust continuum survey \citep{Sch09}, and are distributed in a range of 300\degree $\leq l \leq$ 20\degree and $\vert b \vert \leq 1$\degree. The MALT90 survey simultaneously observed 16 major molecular lines at 90\,GHz, providing a detailed, legacy catalog of molecular line emission for massive star formation clumps \citep{Rat16}. Among the 16 major lines, the $J$=1--0 transitions of HCO$^{+}$ and H$^{13}$CO$^{+}$ lines, which serve as tracers of dense gas \citep{Jac13}, are primarily focused here. 

In addition, the 870\,$\mu$m dust continuum data from the ATLASGAL survey \citep{Sch09}, and the mid-infrared images (3.6, 4.5, 5.6, 8.0, and 24\,$\mu$m) from the {\it Spitzer}-GLIMPSE \citep{Chu09} and -MIPSGAL \citep{Car09} surveys are used to investigate the IR signatures of the star-formation activity of the massive clumps studied here.

\section{results and analysis} \label{sec:results and analysis}
\subsection{Sample refinement}\label{subsec: subsec1}
We applied certain criteria to refine the sample of clumps for robust kinematic analysis (i.e., velocity dispersion, $\sigma$). Firstly, we excluded those clumps located in $\vert l \vert \leq 10\degree$ from the initial set of 3,246 sources observed in the MALT90 survey. These clumps generally have a complex velocity nature that could lead to a highly inaccurate $\sigma$ measurement. 

From the inspection of the clump-averaged H$^{13}$CO$^{+}$~(1--0) spectrum, we eliminated those clumps with multiple velocity components along the line of sight that  can also affect the measurement's accuracy. From the line spectrum, several observed parameters, such as $\sigma$ and the line peak intensity, $I_{\rm p}$, can be measured performing a Gaussian fit to the averaged H$^{13}$CO$^{+}$ spectrum over the effective clump radius catalogued in \cite{Guz15}. Those authors defined the effective radius as $\theta_{\rm eff}=\sqrt{\Omega/\pi}$ where $\Omega$ corresponds to the area with mass surface densities greater than 0.01\,g\,{\pcmsq}.

Based on the measurement of the line peak intensity of \htcop, we discarded those clumps with $I_{\rm p}$ below 3\,rms, where the rms was estimated in a 5.5\,{\vel} wide spectral window free of line emission. As a result, we obtain a refined sample of 249 clumps. Note that a final sample of 221 clumps are used for robust kinematic analysis (see Sect.\,\ref{subsec: subsec2}).
 
\subsection{Clump classification}\label{subsec: subsec2}
Following \citet{Rat16}, we re-examined the evolutionary phases of the massive clumps investigated here, by adopting a classification scheme based on infrared (IR) signatures, as discussed in the literature \egcite{Cha09, Guz15}. To this end, we utilized the {\it Spitzer} images at 3.6, 4.5, 5.8, 8.0, and 24\,$\mu$m. According to \citet{Rat16}, five candidate phases of clumps are determined as follows: quiescent clumps, which are IR-dark in all {\it Spitzer} 3--24\,$\mu$m images; 
2) protostellar clumps, which are associated either with a point-like source indicative of a young stellar object (YSO) at least in one of the {\it Spitzer} images, or with extended 4.5\,$\mu$m emission indicative of the shocks generated by an associated YSO(s);
3) HII region clumps, which are characterized as a compact clumpy 8\,$\mu$m signature typical of polycyclic aromatic hydrocarbon (PAH) emission related to ionized gas from newly born massive stars; 
4) photodissociation region (PDR) clumps, which are dominated by extended 8\,$\mu$m PAH emission typically appearing at the interface between ionized and molecular gas;
5) uncertain clumps, which have ambiguous IR signatures, and thus cannot be grouped into any of the four known phases. In Appendix\,\ref{sec: ApdA}, we provide examples of the classification scheme, showing how different IR signatures between 3.6 and 24\,$\mu$m correspond to the candidate evolutionary phases of clumps. 

As a result, the classification scheme yields five candidate phases of clumps, including 18 quiescent, 100 protostellar, 84 HII region, 19 PDR, and 28 uncertain phases.  
To explore the velocity dispersion as a function of evolution phase (see below), we hereafter will take into account only the 221 known-phase clumps for robust analysis. 
Their  major parameters mentioned above are listed in Table\,\ref{Table} of Appendix\,\ref{sec: ApdB}, including the radius ($R$), mass surface density ($\Sigma$), mass ($M$), luminosity ($L$), and velocity dispersion ($\sigma$) in order from Col.\,5 to Col.\,9.

The bolometric luminosity-mass ratio ($L/M$) is a commonly-used indicator of clump evolution \citep{Mol08}. For the 221 clumps investigated here, their luminosity can be retrieved from \cite{Con17} and \cite{Urq22} who derived it from the spectral energy distribution (SED) fitting to the observed fluxes at least in three wavelength bands between 8\,\um\ and 870\,\um\ \citep{Urq18}.  The calculation of the clump mass is referred to Sect.\,\ref{subsec: subsec3}. 

\begin{figure}
\centering
\includegraphics[width=3.3 in]{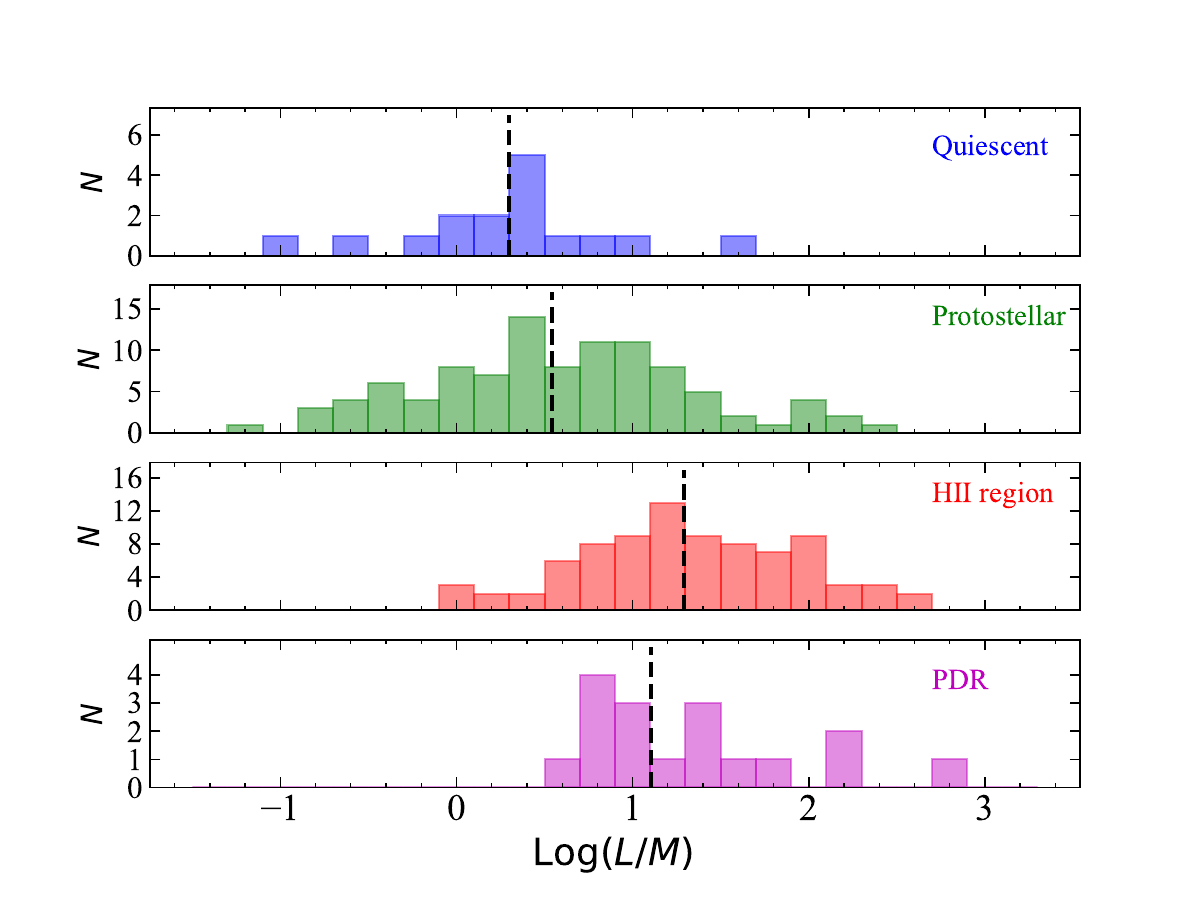} 
\caption{Luminosity-mass ratio distribution for different evolutionary phases of clumps, including the quiescent, protostellar, HII region, and PDR phases. The black vertical dashed line marks the median ratio for each phase. The median value for each phase is 2.0, 3.5, 19.5, 12.8, respectively.}
\label{fig:Luminosity mass ratio distribution}
\end{figure}

Figure\,\ref{fig:Luminosity mass ratio distribution} presents the $L/M$ distribution of the clumps for different evolutionary phases. Note that three clumps (AGAL012.863-00.244, AGAL013.213+00.039, and AGAL014.626-00.562) do not have available luminosity measurements, and thus are not considered in the figure. Overall, the $L/M$ of these clumps increases from the quiescent to HII region phase, as manifested by the median ratio of each phase (see the dashed line in  Figure\,\ref{fig:Luminosity mass ratio distribution}). This result is in good agreement with that of the evolutionary classification from the IR signatures (see above). It is worth noting that there is no significant difference in the $L/M$ distribution between the HII region and PDR clumps. This implies that these two phases of clumps investigated here do not differ significantly from each other. However, they still have the average $L/M$ ratios much higher than the other two phases of clumps, as reflected from the median ratio value of the phases in Fig.\,\ref{fig:Luminosity mass ratio distribution}.

\subsection{Clump mass and mass surface density}\label{subsec: subsec3}
The mass surface density $\Sigma$ and effective radius $R$ of 221 massive clumps investigated here have been estimated by \cite{Guz15}. These parameters were derived from the SED fitting to the photometric fluxes at least in three wavelengths of {\it Herschel} 160, 250, 350, and 500\,$\mu$m, and ATLASGAL 870\,$\mu$m. We calculated here the clump mass as $M = \pi R^2 \Sigma$.

The 221 clumps have 27 to 64,978\,$M_\odot$ for the mass,  and 0.04 to 0.42\,g\,{\pcmsq} for the mass surface density. They have median values of [379, 1061, 1611, 1761]\,$M_\odot$ for the mass, and [0.12, 0.13, 0.12, 0.12]\,g\,{\pcmsq} for the mass surface density, from the quiescent, protostellar, HII region, and PDR phase, respectively. These results suggest that most of the clumps investigated here, if not all, are sufficiently massive and dense for high-mass star formation, as requested by an empirical high-mass star formation threshold of $\Sigma_{\rm crit} \geq 0.05$\,\,g\,{\pcmsq} \egcite{Urq14}.

\subsection{Velocity dispersion}\label{subsec: subsec4}
\begin{figure}
\centering
\includegraphics[width=3.3 in]{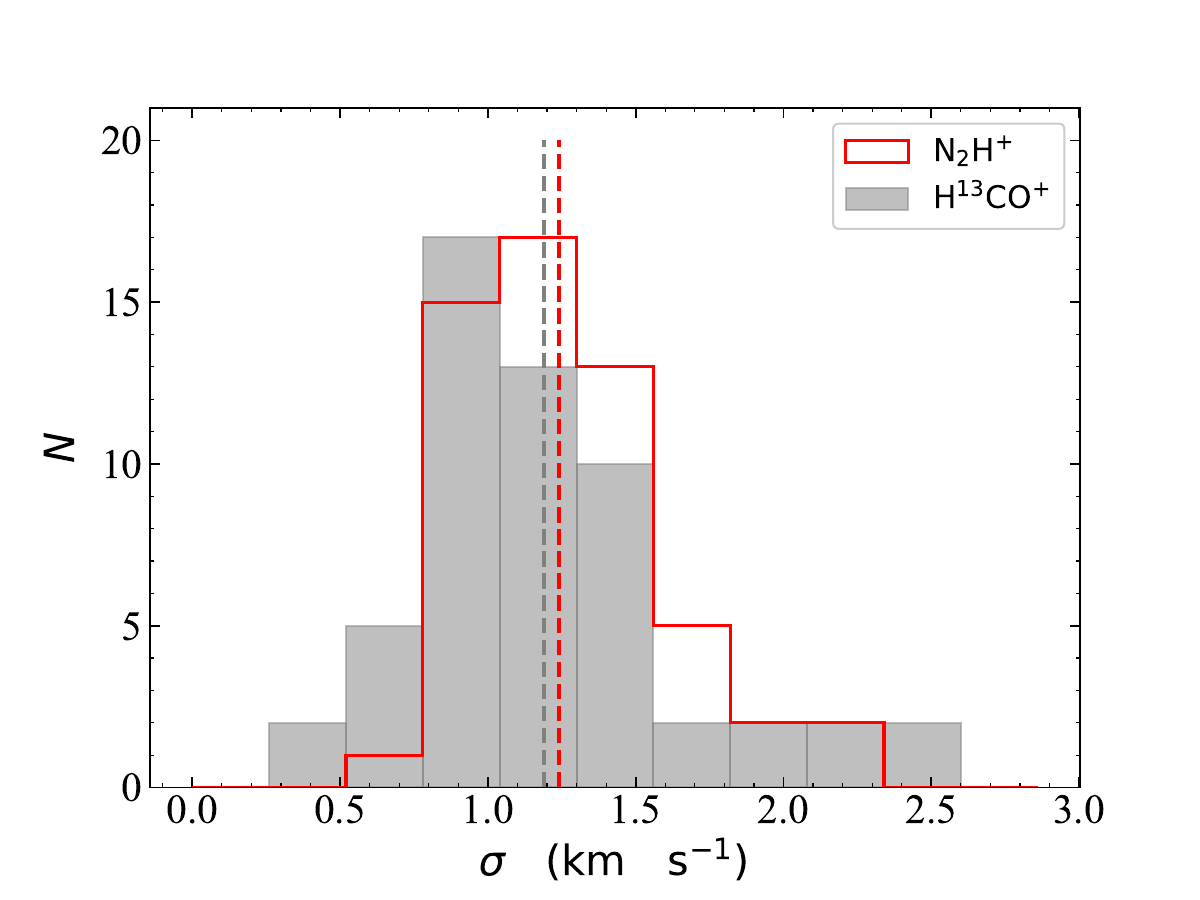} 
\caption{Velocity dispersion distribution of the 55 clumps that overlap between \citet{Tra18} and this work. The gray and red histograms represent the velocity dispersion distributions derived from the {\htcop}~(1--0) line (see Sect.\,\ref{subsec: subsec1}), and the $\mathrm{N_{2}H^{+}}$~(1--0) line \citep{Tra18}, respectively. In turn, the gray and red dashed lines correspond to the mean values of the two distributions, respectively.} 
\label{fig:velocity dispersion distribution compare}
\end{figure}

In our samples, there are 55 clumps that overlap the targets analyzed by \citet{Tra18} using the $\mathrm{N_{2}H^{+}}$~(1--0) line emission. For these clumps, Fig.\,\ref{fig:velocity dispersion distribution compare} displays the comparison of the velocity dispersion distribution between the measurements from two different molecular lines. It turns out that the $\sigma$ measurements from both the {\htcop} (this work) and $\mathrm{N_{2}H^{+}}$ \citep{Tra18} lines present a consistent distribution. Quantitatively, the velocity dispersion distribution obtained from the {\htcop}/$\mathrm{N_{2}H^{+}}$ lines shows median values of 1.2\,{\vel}/1.1\,{\vel}, the mean values of 1.24\,{\vel}/1.19\,{\vel}, and rms values of 0.33\,{\vel}/0.45\,{\vel}.

\begin{figure}
\centering
\includegraphics[width=3.3 in]{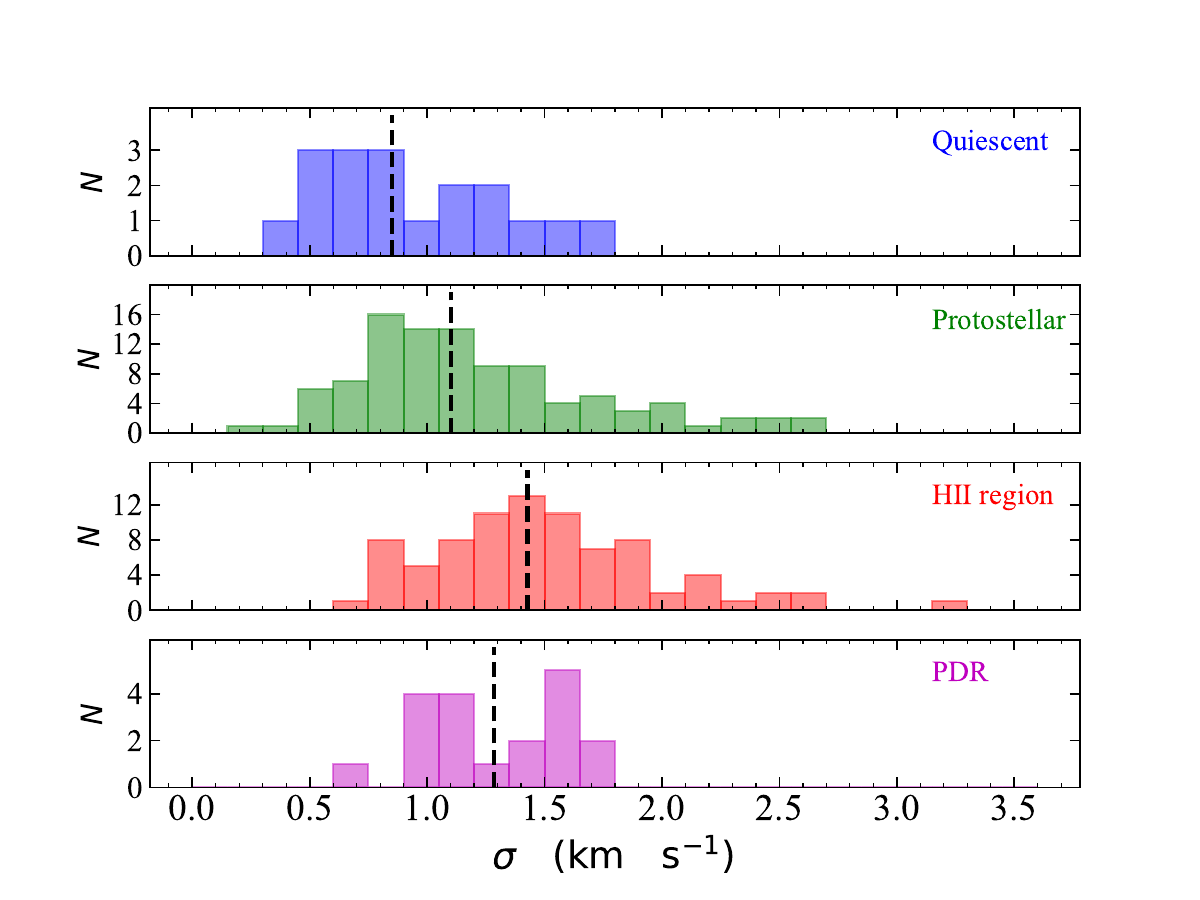} 
\caption{Velocity dispersion distribution for different evolutionary phases of clumps, including the quiescent, protostellar, HII region, and PDR phases. The black vertical dashed line indicates the median ratio for each phase.  The median value for each phase is 0.8\,{\vel}, 1.1\,{\vel}, 1.4\,{\vel}, 1.3\,{\vel}, respectively.
The velocity dispersion increases from the quiescent to HII region phase, but then decreases toward the PDR phase.}
\label{fig:velocity dispersion distribution}
\end{figure}

In Figure\,\ref{fig:velocity dispersion distribution}, we present the velocity dispersion distribution ($\sigma$) of  the four evolutionary phases of clumps.  As can be seen, the velocity dispersion increases distinctively with the evolutionary sequence from the quiescent to HII region phase (see the dashed line for the median value in Fig.\,\ref{fig:velocity dispersion distribution}). This suggests that the velocity dispersion in massive star formation regions can evolve with time, and accordingly could be significantly enhanced by stellar feedback, such as protostellar jets, outflows and the expansion of ionized gas \egcite{Knee2000, Mat02, Qui05, Gol11}. Moreover, the decrease in the velocity dispersion of the PDR clumps can be found compared to the HII-phase counterparts. This could be due to the fact that PDRs are generally located farther away from the central massive star than HII regions, and thus could be less affected by strong stellar feedback. Even so, the velocity dispersion is still higher in the PDR clumps than in the quiescent and protostellar phases, 
which agrees with the result derived from the $L/M$ ratio distribution.

\subsection{Virial parameter}\label{subsec: subsec5}
To determine the gravitational state of clumps, the virial parameter was gauged by {\avir} $= 5\sigma^2R/(GM)$ \citep{Ber1992, Kau13}, where the clump mass, radius, and velocity dispersion ($M$, $R$, and $\sigma$) can be found in Table\,\ref{Table}, and G is the gravitational constant. The critical value of $\alpha_{\rm vir, crit} \simeq 2$ is assumed for the hydrostatic equilibrium of spherical cloud fragments that are not supported by magnetic pressure. Therefore, clumps with {\avir} $ > 2$ could be over-virial, while those with {\avir} $ < 2$ could be sub-virial \citep{Kau13}.

For our sample, the {\avir} values range from 0.11 to 11.65, spanning two orders of magnitude, with a median value of 1.13. Of the 221 clumps, 43 are of {\avir} $ > 2$ while 178 are of {\avir} $ \leq 2$. This suggests that the majority of the clumps are sub-virial provided that they are not supported by magnetic fields. The median {\avir} values are observed to be [1.07, 0.96, 1.35, 0.94] in the quiescent, protostellar, HII region, and PDR phases of clumps, respectively. The protostellar clumps seem to have lower {\avir} values than the quiescent clumps. Note that the quoted median values between both stages of clumps differ only by 0.11, which may not be statistically significant, provided the uncertainties in estimating the masses, sizes, and velocity dispersion. Moreover, the HII region clumps have the highest virial parameters, implying fresh kinetic energy injections into clumps from embedded stellar feedback \citep{Mat02}. Moving onward to the PDR phase,  their  {\avir} values start to decrease, which can be attributed to their lower velocity dispersion than in HII clumps as mentioned in Sect.\,\ref{subsec: subsec4}.

\subsection{Larson's linewidth-size relation}\label{subsec: subsec6}
The Larson's linewidth-size relation, discovered in \citet{Larson1981}, is characterised as a power-law pattern for the velocity dispersion {\sig} and the radius $R$ of MCs,  $\sigma \propto\mathit{R}^{\gamma}$ for $\gamma=0.38$. Later, the index $\gamma$ was modified to be about 0.5 \egcite{Solomon1987,Hey04}. The Larson's $\sigma$--$R$ relation is generally interpreted as a turbulent energy cascade from large to smaller scales in MCs, where turbulence is thought to act against gravitational collapse of MC density structures.

\begin{figure}
\centering
\includegraphics[width=3.3 in]{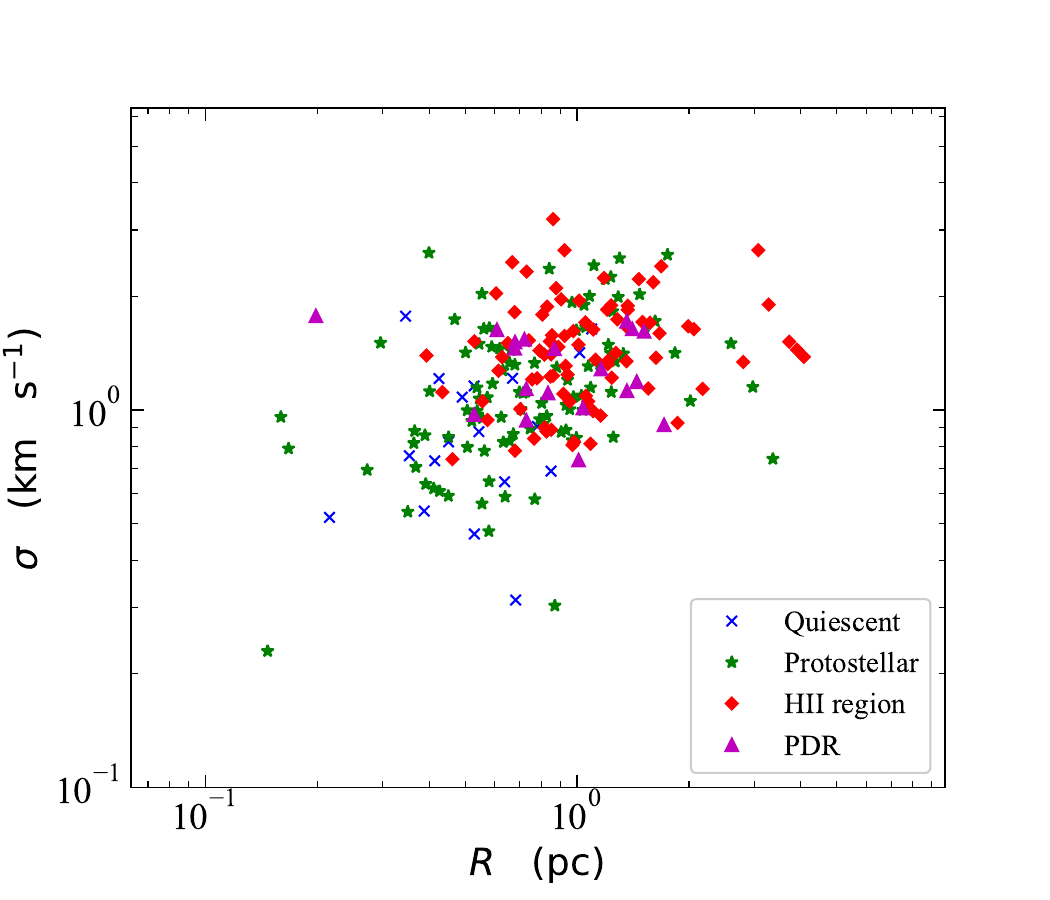} 
\caption{Velocity dispersion-radius relation for 221 massive clumps.}
\label{fig:Velocity dispersion-radius relation}
\end{figure}

Figure\,\ref{fig:Velocity dispersion-radius relation} shows the $\sigma$-$R$ distribution for our sample. We find a weak correlation between the $\sigma$ and $R$ parameters, which agrees with a lower value of the Pearson's correlation coefficient, $\rho = 0.43$. This result does not seem to obey the Larson's $\sigma$-$R$ relation. However, it is worth noting that our sample covers a limited range of radii, and thus the tight $\sigma$-$R$ relation/correlation cannot emerge. \cite{Li23} obtained the similar result for the same analysis in a limited dynamical range of spatial scales. Moreover, if the whole sample is divided into the four phases of clumps (see Figure\,\ref{fig:Velocity dispersion-radius relation}), they will not present any degree of the $\sigma$-$R$ correlation.  We therefore cannot explore how the Larson's $\sigma$-$R$ relation varies with star-formation environment. 

To address the issue of a limited range of spatial scales when investigating the $\sigma$-$R$ relation, we collected the data of the $\sigma$ and $R$ parameters from the literature, including \cite{Heyer09} for a sample of GMCs in the relatively high-density area enclosed within 1/2 maximum isophote of H$_2$ column density, \cite{Tra18} for a sample of massive clumps where the overlapping 55 sources are replaced with ours, \cite{Oha16} for a sample of high-mass star formation clumps and cores where we only keep those with the measurements of both $\sigma$ and $R$, and  \cite{Per06}, \cite{Lu18}, and \cite{Li23}  for additional high-mass star-formation cores. Using the same approach as in Sect.\,\ref{subsec: subsec5}, the {\avir} parameter was estimated for these new added density structures given their mass from the literature listed above.

\begin{figure*}
\centering
\includegraphics[width=6.4 in]{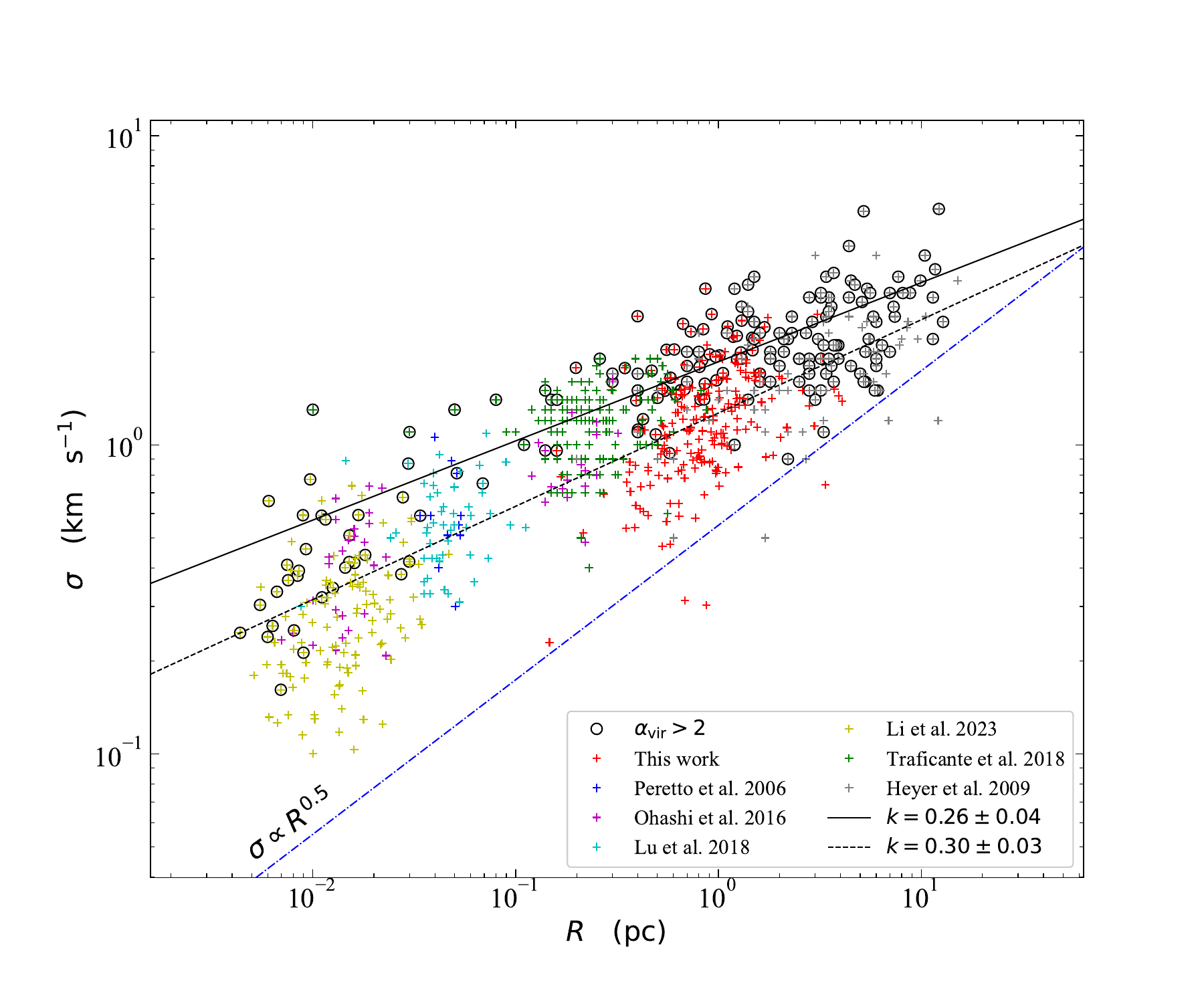} 
\caption{Velocity dispersion-radius relation for different density structures from GMCs to dense cores. The sources with {\avir} $> 2$ are circled in black, while those with {\avir} $< 2$ are not. The black solid line is the linear regression fit to the data points with {\avir} $> 2$, for a best fit slope of $0.26 \pm 0.04$ and a Pearson’s coefficient of $\rho = 0.75$. The black dashed line is the same but for the data points with {\avir} $< 2$, for a best fit slope of $0.30 \pm 0.03$ and a Pearson’s coefficient of $\rho = 0.78$. For comparison, the universal Larsons's relation is represented by the dash-dotted line.}
\label{fig:velocity dispersion versus radius plot}
\end{figure*}

As shown in Figure\,\ref{fig:velocity dispersion versus radius plot}, the new added cloud structures from the literature enable analysis of the  $\sigma$-$R$ relation in three orders of magnitude of spatial scales. We find that the  $\sigma$-$R$ correlation turns tight for all of the cloud structures as characterized by a Pearson's correlation coefficient of $\rho = 0.85$. The corresponding least squares fit in log-log space yields a slope of $\gamma=0.31$. If all of the cloud structures are divided into two categories by the critical $\alpha_{\rm vir, crit} = 2$,  both of those with {\avir} $ > 2$  and with {\avir} $ < 2$ are still tightly correlated in the $\sigma$-$R$ distribution with  $\rho \geq 0.75$. They can be fitted with a power law for a slope of $\gamma = 0.26 \pm 0.04$, and $\gamma = 0.30 \pm 0.03$, respectively. 
Overall, for all the cloud structures as a whole or for two subsamples separated by their {\avir} values, the derived slopes of the $\sigma$-$R$ relation are similar (i.e.,  $\gamma \sim 0.30$), but shallower than the universal Larson's relation. This result suggests that if the slopes were a result of turbulence in molecular cloud structures, the turbulent energy would cascade with a decreasing scale shallower than that reflected from the universal Larson's relation, which could be due to additional inputs of turbulent motions for example by local star-forming feedback (e.g., stellar winds, outflows) especially on smaller scales such as clumps and cores (see Sect.\,4.1). Note that this shallower slope could also be a natural consequence of the cloud collapse since it can increase non-thermal motions of the cloud, which are more structured than random turbulence \egcite{Vaz07, Bal11a, Bal18, Iba16, Iba22}. Disentangling between these two types of consequences requires more dedicated investigations, which are beyond the scope of this work.

\subsection{${\Lcal}$--$\Sigma$ relation}\label{subsec: subsec7}

\begin{figure*}
\centering
\includegraphics[width=6.4 in]{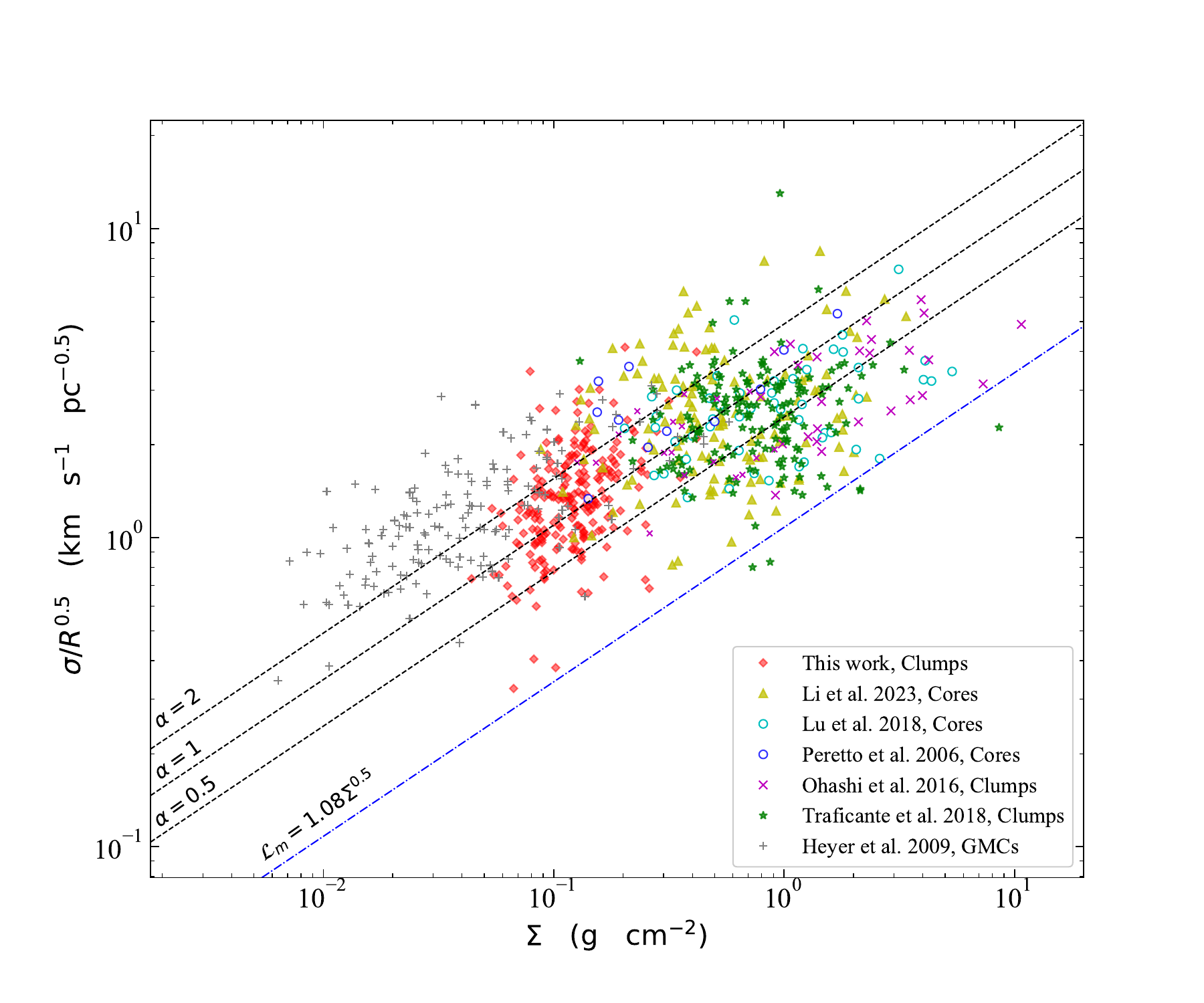} 
\caption{${\Lcal}$--$\Sigma$ relation for different density structures from GMCs to dense cores, where ${\Lcal}$ is the {\sig}$/ \mathit{R}^{0.5}$ ratio.  The black dashed lines correspond to different predicted ${\Lcal}$--$\Sigma$ relation at the virial parameters of 0.5, 1, and 2. The dash-dotted line indicates the predicted, magnetic version of the ${\Lcal}$--$\Sigma$ relation, ${\Lcal}_m=1.08\ \Sigma^{0.5}$, which corresponds to a simple model of magnetically
supported clouds in an equilibrium state.}
\label{fig:Heyer plot}
\end{figure*}

The dynamics of MCs can be regulated by global gravitational collapse as inferred from the ${\Lcal}$--$\Sigma$ scaling relation that follows ${\Lcal}$$=\sigma/R^{0.5} \propto \Sigma^{0.5}$ \citep{Heyer09}.
We inspected the ${\Lcal}$--$\Sigma$ relation only for our sample of massive clumps (not shown here), and found no strong correlation (the Pearson's coefficient $\rho = 0.42$) due to the limited dynamical range of the $\Sigma$ parameter. Therefore,  we plot in Fig.\,\ref{fig:Heyer plot} the ${\Lcal}$--$\Sigma$ relation with a larger dynamical range of $\Sigma$ by including other cloud structures (see Sect.\,\ref{subsec: subsec6}) in addition to our sample. As a result, a strong correlation between ${\Lcal}$ and $\Sigma$ in the log-log space can be found  with the Pearson's coefficient $\rho = 0.83$. 

We find an increasing trend of the ${\Lcal}$ ratio with $\Sigma$. This  trend corresponds to the transition from the cloud structures of {\avir} $ > 2$ to those of {\avir} $ < 2$ with the former having lower ${\Lcal}$ ratios than the latter. Here the former structures contain mostly (64.6\%) large-scale clouds while the latter consist mostly (85\%) of small-scale clumps and cores. Quantitatively, the median value of the virial parameter is 2.69 for the GMCs, and 0.86 for the clumps and cores, suggesting the decrease of the virial parameter from large-scale GMCs to smaller-scale clumps and cores. The surface density increases from giant molecular clouds to dense cores, with a median value of surface density being 0.04\,g\,cm$^{-2}$ for the GMCs in contrast to 0.38\,g\,cm$^{-2}$ for the clumps and cores. To conclude, the different behaviors between these two types of density structures in the ${\Lcal}$ distribution suggest a dynamical decoupling between the large-scale clouds and the smaller-scale clumps/cores, which agrees with the finding of \citet{Per23}.

\section{Discussion}\label{sec:discussion}
\subsection{Relation of cloud dynamics to turbulence, gravity, and magnetic field }\label{Dis subsec1}

Turbulence is ubiquitous in molecular clouds. The turbulence on large scale can be driven by various sources such as cloud-cloud collisions, supernova explosions, and the expansion of HII regions \egcite{Ost11,2013ApJ...774L..31I}. As shown in Fig.\,\ref{fig:velocity dispersion versus radius plot}, strong turbulent motions appear in the large-scale MCs, as reflected in the high velocity dispersion at scales of $\sim$10\,pc. This implies that turbulence would have a relatively stronger effect on the large-scale dynamics of molecular clouds and thus delay their gravity collapse. Nevertheless, the turbulence cascades toward small scales in MCs, resulting in turbulence gradually becoming less important. Notably, in small-scale dense cores with a size of $\sim 0.01$\,pc, the gas velocity dispersion sharply decreases. This implies that if  there are no internal, localized driving sources to maintain it, turbulence will quickly dissipate in dense cores.

\begin{figure*}
\centering
\includegraphics[width=6.4 in]{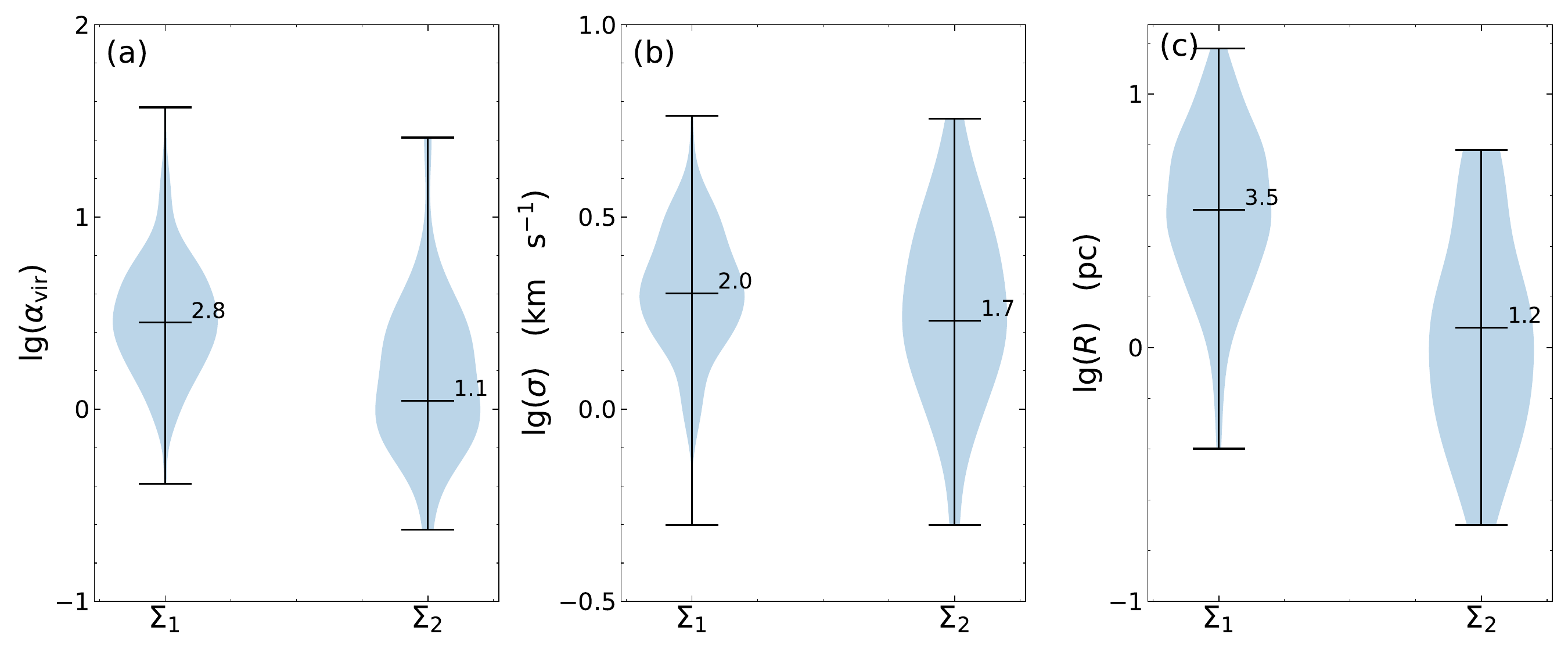} 
\caption{Violin plot of the $\alpha_\mathrm{vir}$, velocity dispersion, and size distribution for the GMCs in two sub-samples  with different mass surface density. The shape of each distribution shows the probability density of the $\alpha_\mathrm{vir}$, velocity dispersion, and size, respectively. The black horizontal bars from the top to bottom in violin plot represent the maximum, median, and minimum values, respectively. The $\Sigma_1$ on the horizontal axis represents the median of sub-samples with low mass surface density, while $\Sigma_2$ represents the median of sub-samples with high mass surface density, which are 0.03\,g\,cm$^{-2}$ and 0.16\,g\,cm$^{-2}$, respectively.}
\label{fig:heyer gmc}
\end{figure*}

In order to further investigate the interaction between turbulence and gravity, we conducted a more detailed analysis of the GMC sample from \cite{Heyer09}. We divided the sample into low surface density sub-samples, which contained 129 sources ($\Sigma < 0.1$\,g\,cm$^{-2}$), and high surface density sub-samples, containing 29 sources ($\Sigma \geq 0.1$\,g\,cm$^{-2}$). The difference in median surface density between two sub-samples is about five factor, with values of 0.03\,g\,cm$^{-2}$ and 0.16\,g\,cm$^{-2}$, respectively. As shown in Figure\,\ref{fig:heyer gmc}, an observed trend indicates that as the surface density increases, the size of clouds correspondingly decreases, reflecting cloud contraction. Furthermore, the evolution from an over-virial state to a sub-virial state during cloud contraction suggests that gravity becomes relatively more important compared to turbulence. In addition, although the turbulence dissipates during cloud contraction, the variation in velocity dispersion of MCs is not significant, indicating that the gravitational contraction contributes to non-thermal motions but not significantly. Such gravity-driven non-thermal motions are plausible on large-scale MCs, as several analysis of the column density distribution function of large-scale clouds have indicated signs of global collapse \egcite{Sch13, Sch15}.

\begin{figure}
\centering
\includegraphics[width=3.3 in]{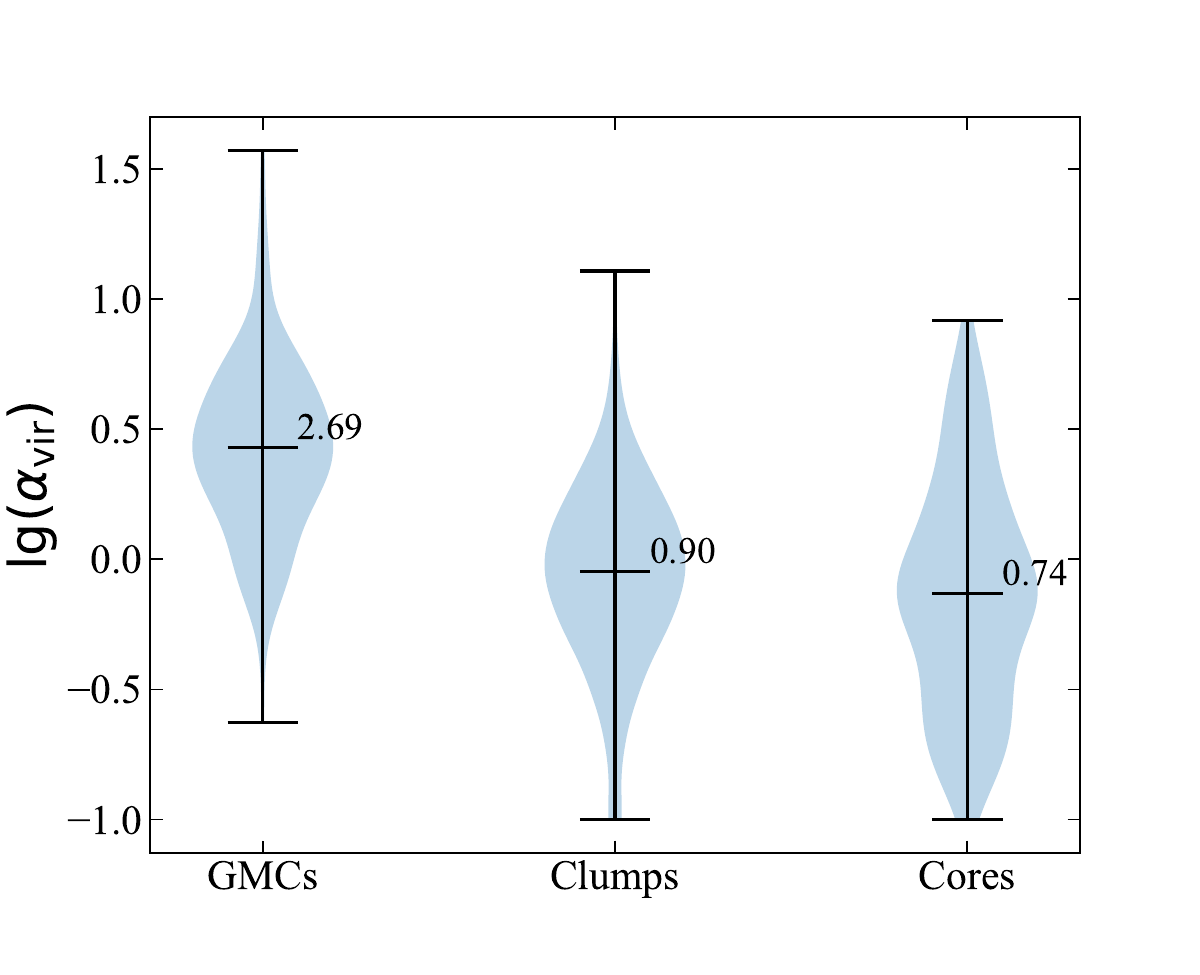} 
\caption{Violin plot of the {\avir} distribution for different density structures. The shape of each distribution shows the probability density of the {\avir}. The black horizontal bars from the top to bottom in violin plot represent the maximum, median, and minimum values, respectively. For different density structures, the median values of the {\avir} are 2.69, 0.90, and 0.74, respectively. The data for GMCs is sourced from \cite{Heyer09}. The data for clumps are from \cite{Oha16}, \cite{Tra18}, and this work, respectively. The data for cores are from \cite{Per06}, \cite{Oha16}, \cite{Lu18}, and \cite{Li23}, respectively.}
\label{fig6}
\end{figure}

We can further explore how turbulence and gravity affect the cloud dynamics by scrutinizing the observed ${\Lcal}$--$\Sigma$ relation in Figure \,\ref{fig:Heyer plot} from left to right. The relation shows that clouds change from an over-virial state (i.e., $\alpha_\mathrm{vir} > 2$ for the majority of large-scale clouds) to a sub-virial state ($\alpha_\mathrm{vir} < 2$ for most smaller-scale density strucrures) as they become  smaller and denser, which is also evident in the distribution of the virial parameter across density scales in Figure\,\ref{fig6}. This observed transition suggests that the effect of turbulence on the dynamics of clouds could gradually weaken during their evolution, while the effect of gravity gradually strengthen. This finding is consistent with the results of \cite{Per23}, who found that small-scale, dense density structures up to filaments are governed by gravity over turbulence by  performing systematical, multiscale analysis of the virial parameter and velocity dispersion of 27 filamentary clouds. Moreover, the simulation of dense core collapse suggests that turbulence does not have a significant impact on the dynamics of dense cores if it is dissipated promptly after being generated \citep{Gue20}.

Therefore, we suggest that both turbulence and gravity could play crucial roles in regulating together the dynamics of molecular clouds. The ubiquitous turbulence would impact profoundly the dynamics of large-scale MCs before they become sub-virial; subsequently, the self-gravity in sub-virial MCs becomes important down to core scales, leading to global hierarchical collapse as proposed in \citealt{Vaz19}.

It is worth noting that large-scale clouds are generally observed to transition from an over-virial state to a sub-virial state (see Fig.\,7), subsequently fragmenting hierarchically into smaller, sub-virial density structures such as cores and clumps. This fragmentation aligns with the GHC model, where gravity dominates the collapse of molecular clouds \citep{Har01, Bal11a, Bal18, Vaz19}. 
Interestingly, some smaller structures, especially at the core and clump scales, appear to be in an over-virial state (Fig.\,\ref{fig:velocity dispersion versus radius plot}). As predicted in the GHC model, the velocity dispersion in these structures could be amplified by non-thermal motions induced by cloud collapse. These motions, being a result of an ordered collapse process, could be more structured than random turbulence, leading to a pseudo over-virial state. 
On the other hand, these particular small-scale structures could be really over-virial if they are subjected to strong local stellar feedback, such as outflows and stellar winds \citep{Pan23}.

Strong magnetic fields in MCs can counteract the effects of turbulence and gravity, as evidenced by the ordered appearance of field morphology from thermal dust emission polarization \egcite{Li13, Pla16, Sol16, Gu19}. In the presence of strong magnetic fields, cloud motions become sub-Alfv\'enic, meaning that neutral gas is dynamically coupled to the field through ion-neutral collisions. These cloud motions may originate from the propagation of large amplitude, long wavelength Alfv\'en waves \citep{Heyer09}. \cite{Mou06} proposed a simple model of magnetically supported clouds that can reach an equilibrium state when the cloud surface density equals the magnetic critical surface density, given by the equation $\Sigma_{\rm crit, m} = B/(63G)^{1/2}$. If the magnetic field strength can be predicted from the observed $\sigma^2/R$ ratio using the equation  $B \sim (45/G)^{1/2} \sigma^2/R$, as demonstrated by \cite{Mye1988} for several nearby clouds with thermal Zeeman OH measurements, then magnetically supported clouds in equilibrium will satisfy the magnetic version of the ${\Lcal}$--$\Sigma$ relation, ${\Lcal}_m=1.08\ \Sigma^{0.5}$. Figure 6 shows that the predicted  ${\Lcal}_m$--$\Sigma$ relation is much lower than the observed data points. This suggests that the magnetic fields alone may not be sufficient to maintain cloud against cloud gravitational collapse \egcite{Tra18}. Nevertheless, magnetic fields may still have a crucial role in other aspects of cloud dynamics, such as filament formation and protostellar disk support. To fully understand the role of magnetic fields in cloud's dynamics, we require a comprehensive approach that accounts for the complex interplay between turbulence, gravity, and magnetic fields on different scales.

It is important to note that the discussions we made above on the {\sig}-$R$ and ${\Lcal}$--$\Sigma$ relations may be affected by observational biases, as the data were obtained using different telescopes, molecular tracers and analysis techniques. In addition, one must recognise that systematics from different surveys may influence current analysis. For example, we excluded the CO cloud data from \citet{Rom10}, who gathered data from both the University of Massachusetts-Stony Brook (UMSB) and Galactic Ring surveys. Our data from \citet{Heyer09}, however, solely originate from the latter. The Roman-Duval's data, leading to virial parameters averaging six times lower than Heyer's (as depicted in Fig.\,1 of \cite{Kau13}), does not conform to the trend shown in Fig.\,\ref{fig6}.Besides, different molecular tracers can have different physical properties and trace different density regions. Therefore, for more robust discussions on the {\sig}-$R$ and ${\Lcal}$--$\Sigma$ relations, we require in the future consistent, multiscale kinematic observations of molecular clouds using the same telescope and molecular tracer as much as possible.

\subsection{Implication on high-mass star formation}\label{Dis subsec2}
As discussed earlier, the  $\sigma$--$R$ relation could provide insight into the role of turbulence and gravity in molecular cloud dynamics. This would allow us from a global view to place a constraint on the latest theoretical models of high-mass star formation, such as GHC and I2. Both models suggest that high-mass star formation in molecular clouds is a multiscale process involving fragmentation and mass accretion at various density scales. The strong correlation observed in the $\sigma$--$R$ relation supports the kinematic connection between different density structures from clouds to cores, and thus the multiscale nature of high-mass star formation.

Mass accretion is one of the key multiscale physical processes. Both GHC and I2 models agree that gravity drives mass accretion on small scales, such as cores/clumps and even filaments. However, they differ in their predictions for larger scales. That is, the GHC model posits that gravity primarily drives hierarchical mass accretion across all density scales of a cloud \citep{Vaz19}. Conversely, the I2 model proposes that turbulence regulates mass inflow and accretion on large-scale clouds \citep{Pad20}, with the self-gravity of the clouds assuming control at smaller density scales.
As discussed in Sect.\,\ref{Dis subsec1},  the $\sigma$--$R$ and ${\Lcal}$--$\Sigma$ relations we observed indicate that both turbulence and gravity can control the dynamics of MCs, with the former possibly dominating in large-scale over-virial state clouds and the latter  dominating in sub-virial density structures including large-scale clouds down to smaller-scale  clumps and cores. This finding implies that the GHC model effectively accounts for high-mass star formation in sub-virial clouds, which form smaller hierarchical density structures due to gravitational instability. However, our analysis also argues that the role of turbulence may not be overlooked until large-scale clouds reach a sub-virial state. This argument is consistent with the finding of the ubiquitous filamentary density structures in HI clouds where the self-gravity of clouds is not as important as turbulence \egcite{HH23,Liu24}.

\section{Conclusions}\label{sec:Conclusions}
We have investigated the implication of the velocity dispersion scalings on high-mass star formation in molecular clouds, including the scalings of Larson's linewidth--size ($\sigma$--$R$) and Larson's ratio--mass surface density ($\Lcal$--$\Sigma$; here ${\Lcal}$$=\sigma/R^{0.5}$). Our investigations were based on systematic analysis of the $\sigma$ parameter of well-selected 221 massive clumps, complemented with published samples of other hierarchical density structures of molecular clouds over spatial scales of 0.01--10\,pc. Our investigations would allow to put a constraint to theoretical  models of multiscale, dynamical high-mass star formation.  Our major results can be summarized as follows:

\begin{itemize}

     \item The high median value of mass surface density of $>0.05$\,g\pcmsq\ indicates that the majority of the 221 massive clumps are most likely to form high-mass stars. Based on the IR signatures between 3.6 and 24\,\um, these clumps are classified into four evolutionary phases, including 18 quiescent, 100 protostellar, 84 HII region, and 19 PDR clumps. 
     \item The velocity dispersion ($\sigma$) of the clumps increases with the evolutionary sequence from the quiescent to HII region phase, which is related to enhanced stellar feedback in more evolved phases. Then, the $\sigma$ distribution decreases toward the PDR clumps, but is still higher than in the quiescent and protsellar phases. This could be because PDRs are less influenced by stellar feedback than HII regions.
     \item In the relation of Larson's linewidth--szie ($\sigma$--$R$) for our sample alone, we find a weak correlation (the Pearson’s correlation coefficient $\rho=0.43$) due to the limited range of spatial scales of the sample. In the relation of Larson's ratio--mass surface density ($\Lcal$--$\Sigma$), we find the same result ($\rho=0.42$) but due to the limited range of the mass surface density of the sample. 
     \item Both the $\sigma$--$R$ and $\Lcal$--$\Sigma$ relations show strong correlations ($\rho>0.75$) when combing our sample with other density structures collected from the literature, including  molecular clouds, massive clumps, and massive cores. 
     \item In the $\sigma$--$R$ relation, the strong correlation  suggests that if turbulence is important in regulating the dynamics of cloud structures, the turbulent motions will cascade with a decreasing scale. 

    \item In the $\Lcal$--$\Sigma$ relation, clouds transition from over-virial state ($\alpha_\mathrm{vir} > 2$) to sub-virial state ($\alpha_\mathrm{vir} < 2$) as they become smaller and denser. This could reflect a shift in the governing force from turbulence to gravity. 
\end{itemize}

    The results inferred from both $\sigma$--$R$ and $\Lcal$--$\Sigma$ relations for different density structures from large-scale MCs to smaller-scale filaments, clumps, and dense cores, suggest that both turbulence and gravity could be important in regulating together the dynamics of hierarchical density structures of molecular clouds, with turbulence possibly prevailing in large-scale over-virial state clouds and gravity prevailing in sub-virial state clouds and smaller-scale hierarchical density structures such as clumps and cores.
    We suggest that the multiscale physical process of high-mass star formation hinges on the self-gravity of sub-virial molecular clouds (MCs). This perspective aligns with the framework of the GHC model  proposed by \cite{Vaz19}. However, we emphasize that the influence of turbulence may not be dismissed until large-scale clouds attain a sub-virial state, which is yet to be confirmed by future multiscale kinematic observations of molecular clouds using the same telescope and molecular tracer as much as possible.

\vspace{5mm}
\noindent
We thank the anonymous referee for comments and suggestions that greatly improved the quality of this paper. This work has been supported by the National Key R$\&$D Program of China (No.\,2022YFA1603101). H.-L. Liu is supported by National Natural Science Foundation of China (NSFC) through the grant No.\,12103045, and , and by Yunnan Fundamental Research Project (grant No.\,202301AT070118, 202401AS070121). S.-L. Qin is supported by NSFC under No.\,12033005.

\appendix
\section{Examples of classification}\label{sec: ApdA}
\newpage
\begin{figure*}[h]
\centering
\setlength{\abovecaptionskip}{0. in}
\includegraphics[width=6.3 in]{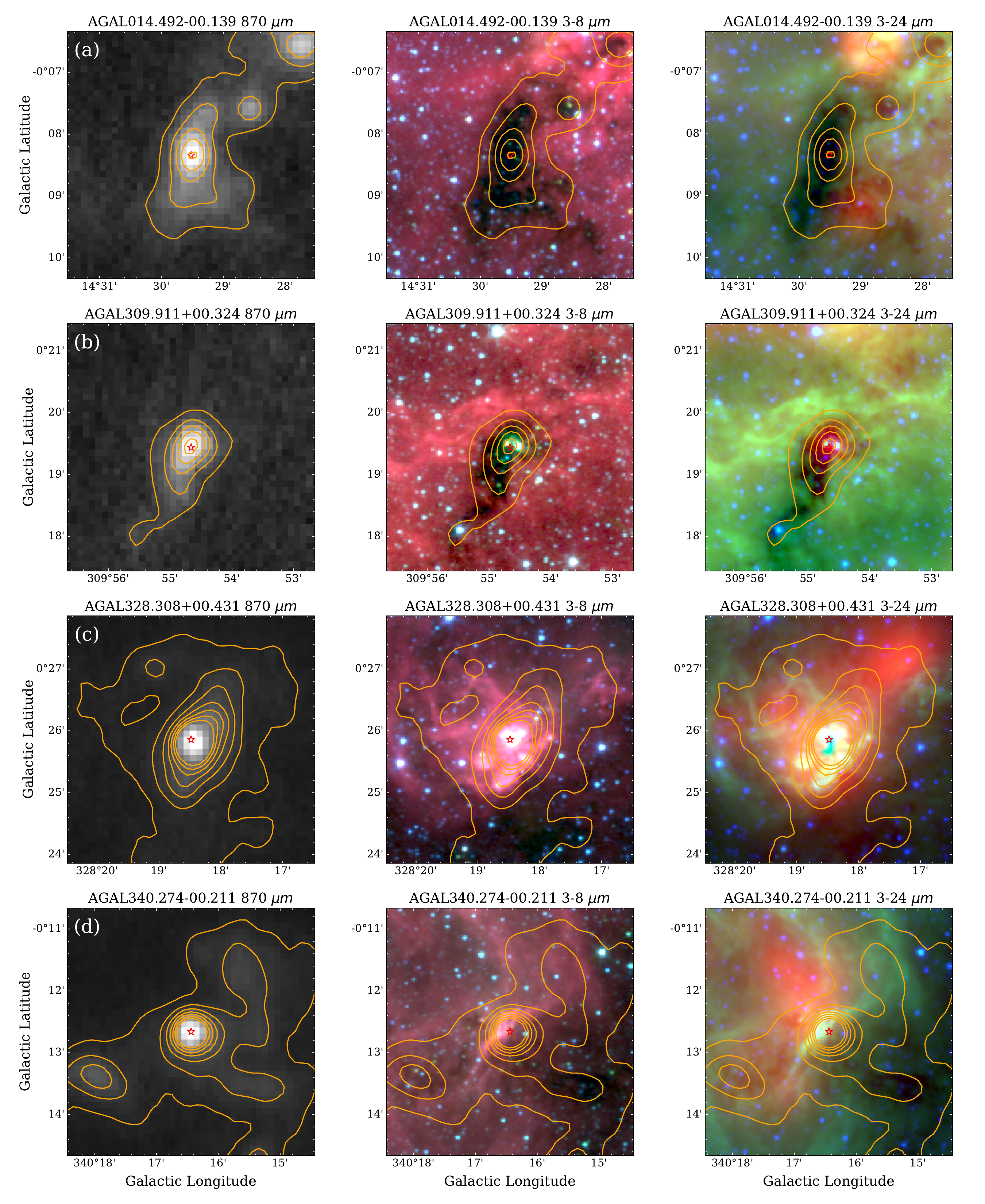} 
\caption{An example of clump classification inferred from {\it Spitzer} images overlaid with 870\,\um\ dust continuum emission. In each row from left to right, the images are 
ATLASGAL 870\,{\um} dust continuum emission in grey-scale, {\it Spitzer} three-colour image of 3.6, 4.5 and 8.0\,{\um}, and {\it Spitzer} three-color image of 3.6, 8.0 and 24\,{\um}. the yellow contours indicate the 870\,{\um} dust continuum emission, starting from 0.3 Jy/beam with a step of 0.5 Jy/beam. Rows (a)--(d) correspond to the examples of 
 classified phases of clumps from the quiescent, protostellar, HII region, and PDR phase, respectively. In all images,  the clump name is indicated at the top of each panel. 
The red star symbol identifies peak emission at 870\,\um\ of the clump. }
\label{fig:three color map}
\end{figure*}

\section{Physical properties of clumps}\label{sec: ApdB}
\startlongtable
\begin{deluxetable*}{ccccccccccc}\label{Table}
\tabletypesize{\footnotesize}
\tablecaption{Physical properties of clumps} 
\tablehead{
\colhead{ID} & \colhead{Sname} & \colhead{$l$} & \colhead{$b$} & \colhead{$\mathit{R}$} & \colhead{$\mathit{\Sigma}$} & \colhead{$\mathit{M}$} & \colhead{$\mathit{L}$} & \colhead{$\sigma$} & \colhead{$\alpha_{\rm{vir}}$} & \colhead{evol.phase}  \\
       &  & \colhead{deg} & \colhead{deg}  & \colhead{pc} & \colhead{g cm$^{-2}$} & \colhead{$M_{\odot}$} & \colhead{$L_{\odot}$} & \colhead{km s$^{-1}$} &   &     }
\startdata 
 1   &  AGAL010.168-00.362   &   10.168   &  -0.362   &    0.86  $\pm$      0.17   &   0.08  $\pm$     0.01   &    883  $\pm$     260    &   231739   &    3.2  $\pm$     0.14   &  11.65  $\pm$     4.43   &     HII region  \\ 
 2   &  AGAL010.204-00.347   &   10.204   &  -0.347   &    0.68  $\pm$      0.14   &   0.18  $\pm$     0.01   &   1263  $\pm$     380    &    29376   &   1.46  $\pm$      0.1   &   1.33  $\pm$      0.5   &            PDR  \\
 3   &  AGAL010.213-00.322   &   10.213   &  -0.322   &    1.11  $\pm$      0.22   &   0.14  $\pm$     0.02   &   2684  $\pm$     810    &    40272   &   2.42  $\pm$     0.13   &    2.8  $\pm$     1.07   &   Protostellar  \\
 4   &  AGAL010.228-00.207   &   10.228   &  -0.207   &     1.0  $\pm$       0.2   &   0.13  $\pm$     0.01   &   1968  $\pm$     590    &     2979   &   0.84  $\pm$     0.09   &   0.42  $\pm$     0.16   &   Protostellar  \\
 5   &  AGAL010.299-00.147   &   10.299   &  -0.147   &    0.93  $\pm$      0.19   &   0.13  $\pm$     0.01   &   1622  $\pm$     490    &    98628   &   2.65  $\pm$     0.18   &   4.65  $\pm$     1.77   &     HII region  \\
 6   &  AGAL010.323-00.161   &   10.323   &  -0.161   &    0.98  $\pm$       0.2   &    0.1  $\pm$     0.01   &   1465  $\pm$     440    &   179473   &   1.62  $\pm$     0.14   &   2.03  $\pm$     0.77   &     HII region  \\
 7   &  AGAL010.342-00.142   &   10.342   &  -0.142   &    0.59  $\pm$      0.12   &   0.13  $\pm$     0.01   &    690  $\pm$     210    &    11695   &   1.17  $\pm$     0.15   &   1.38  $\pm$     0.52   &   Protostellar  \\
 8   &  AGAL010.472+00.027   &   10.472   &   0.027   &    3.07  $\pm$      0.61   &   0.13  $\pm$     0.01   &  18340  $\pm$    5500    &   442588   &   2.65  $\pm$     0.21   &   1.37  $\pm$     0.52   &     HII region  \\
\enddata

\tablecomments{Physical properties of the 221 clumps. Col.2-4: Clump name and coordinates; Col.5-7: Clump radius, average mass surface density, and mass; Col.8: Clump luminosity; Col.9-10: Clump velocity dispersion and virial parameter; Col.11: the evolutionary phase of clumps. Three dots in the clump luminosity column signify that the luminosity data for these clumps was not collected. Note that the clump radius and average mass surface density are retrieved from \cite{Guz15}, while the clump luminosity is obtained from \cite{Con17} and \cite{Urq22}. The remaining derived parameters are carried out in this work.}
\end{deluxetable*}

\bibliography{AJ}{}
\bibliographystyle{aasjournal}

\end{document}